\documentclass[pdftext,a4wide,12pt]{article}
\usepackage[dvips]{graphicx}
\usepackage{epsfig}
\usepackage{amsfonts}
\usepackage{graphics,color}
\begin{document}
\def\ti{\tilde}
\def\da{\dagger}
\def\a{\alpha}
\def\b{\beta}
\def\g{\gamma}
\def\l{\lambda}
\def\d{\delta}
\def\k{\kappa}
\def\p{\partial}
\def\t{\theta}
\def\s{\sigma}
\def\G{\Gamma}
\def\ap{\approx}
\def\v{\varepsilon}
\def\tr{\textrm}
\def\nn{\nonumber}
\def\w{\wedge}
\def\hw{\hat{\omega}}
\def\K{K\H{a}hler}
\def\o{\omega}
\def\na{\nabla}
\def\t{\theta}
\def\s{\sigma}
\def\G{\Gamma}
\def\D{\Delta}
\def\v{\varepsilon}
\def\tr{\textrm}
\def\nn{\nonumber}
\def\w{\wedge}
\def\hw{\hat{\omega}}
\def\K{K\H{a}hler}
\def\o{\omega}
\newcommand{\be}{\begin{eqnarray}}
\newcommand{\ee}{\end{eqnarray}}
\newcommand{\beq}{\begin{equation}}
\newcommand{\eeq}{\end{equation}}
\newcommand{\rc}{\nonumber\\}

\pagenumbering{arabic}
\renewcommand{\theequation}{\thesection.\arabic{equation}}

\definecolor{M_Beige}         {rgb}{0.96 , 0.96 , 0.86}
\newcommand{\CBei}[1]{\textcolor{M_Beige}{#1}}
\definecolor{M_Brown}         {rgb}{0.65 , 0.16 , 0.16}
\definecolor{M_Gold}          {rgb}{0.12 , 0.84 , 0.30}
\newcommand{\CGol}[1]{\textcolor{M_Gold}{#1}}
\definecolor{M_LemonChiffon}  {rgb}{1.00 , 0.98 , 0.80}
\newcommand{\CLem}[1]{\textcolor{M_LemonChiffon}{#1}}
\definecolor{M_Orange}        {rgb}{1.00 , 0.60 , 0.00}
\newcommand{\Ccya}[1]{\textcolor{M_Orange}{#1}}
\definecolor{M_Pink}          {rgb}{1.00 , 0.75 , 0.80}
\newcommand{\CPin}[1]{\textcolor{M_Pink}{#1}}
\definecolor{M_Gre}          {rgb}{0.05 ,0.46 , 0.00}
\newcommand{\Cred}[1]{\textcolor{M_Gre}{#1}}

\newcommand{\Cbla}[1]{\textcolor{black}{#1}}
\newcommand{\Cblu}[1]{\textcolor{blue}{#1}}
\newcommand{\Cgre}[1]{\textcolor{red}{#1}}
\newcommand{\Cyel}[1]{\textcolor{yellow}{#1}}
\newcommand{\Cwhi}[1]{\textcolor{white}{#1}}

\begin{titlepage}
\begin{center} \Large{ \bf On the M-theory description of supersymmetric gluodynamics}


\end{center}

\vskip 0.3truein
\begin{center}
Felipe Canoura${}$
\footnote{canoura@fpaxp1.usc.es}
and
Paolo Merlatti${}$
\footnote{merlatti@fpaxp1.usc.es}

\vspace{0.3in}

${}$Departamento de F\'\i sica de Part\'\i culas, Universidade de
Santiago de Compostela \\and\\
Instituto Galego de F\'\i sica de Altas Enerx\'\i as (IGFAE)\\
E-15782 Santiago de Compostela, Spain

\end{center}
\vskip 1
truein

\begin{center}
\bf ABSTRACT
\end{center}

\noindent We study the stringy description of ${\cal N}=1$ supersymmetric $SU(N)$ gauge theory on ${\mathbb R}^{1,2}\times S^1$. Our description is based on the known Klebanov-Strassler and Maldacena-N\'u\~nez solutions, properly modified to account for the compact dimension. The presence of this circle turns out to be a non trivial modification and it leads us to consider the up-lifted eleven dimensional solution. We discuss some of its properties. Perhaps the most interesting one is that extra BPS M-branes are present. These generate a non-perturbative superpotential that we explicitly compute. Our findings, besides their interest in the gauge-string correspondence, may also have applications in the cosmological KKLT and KKLMMT scenarios.

\vskip4.6truecm
\leftline{US-FT-2/07}
\leftline{hep-th/yymmnnn \hfill March 2007}
\smallskip
\end{titlepage}
\setcounter{footnote}{0}

\tableofcontents

\section{Introduction}

The duality between string and field theory, in the framework originated by the AdS/CFT correspondence \cite{Maldacena:1997re,Gubser:1998bc,Witten:1998qj}, provides powerful tools to investigate the strong coupling dynamics of the latter. The original formulation of the correspondence concerns an ${\cal N}=4$ supersymmetric field theory in four dimensions. Lowering this high degree of symmetry has been the goal of much recent research (see \cite{Aharony:2002up} for some nice reviews on the subject). Besides obvious phenomenological reasons, reducing the symmetry is also important because it gives a better understanding of the duality when it is applied to the description of truly dynamical processes, going beyond the strong constraints imposed by the symmetries.

In view of their rich and quite well understood dynamics, ${\cal N}=1$ supersymmetric Yang-Mills (SYM) theories provide perhaps the best example to be studied in this perspective. One of their nice properties is that, sometimes, their infrared strong coupling properties can be encoded in a superpotential sourced by non-perturbative effects. The non-perturbative generation of a superpotential makes them appealing also for cosmological purposes and the related moduli stabilization problem \cite{Giddings:2001yu,Kachru:2003aw,Kachru:2003sx}. 
It is in general important to find examples where such superpotential can be directly calculated.

In this paper we find such a concrete example by studying the string dual of ${\cal N}=1$ SYM on the cylinder (by cylinder we mean the flat space ${\mathbb R}^{1,2}\times S^1$, {\it i.e.} four-dimensional Minkowski space with one spatial direction compactified). Besides improving the understanding of the string theoretic description of non-perturbative gauge phenomena, this cylindrical geometry may have interesting cosmological applications.
From the field theory point of view, the generation of a superpotential in this geometry is nicely described  in \cite{Davies:1999uw}. 

There are two well known ${\cal N}=1$ dual supergravity solutions we can start from. They are the Klebanov-Strassler (KS) \cite{Klebanov:2000hb} and the Maldacena-N\'u\~nez (MN) \cite{Maldacena:2000yy} ones, respectively related to the field theories living on the worldvolume of fractional and wrapped D$p$-branes. Our analysis can start from both solutions. We decide to describe more carefully how things work in the MN set-up. In that case the configurations sourcing the superpotential are wrapped branes and their microscopic description, at least in certain limits, is better understood than the one of fractional branes in a conifold background. However, we sketch also how things work in the KS case, which is more relevant in the cosmological set-up \cite{Kachru:2003aw}. 
 
The MN solution describes the infrared (IR) of ${\cal N}=1$ pure SYM theory. Its ultraviolet (UV) completion is instead related to little string theory and the two regimes of the theory are not smoothly connected in terms of a unique solution (they are S-dual to each other). The source of this problem is the bad asymptotic behavior of the dilaton. 
On the gauge theory side this reflects the difficulties of joining the weak coupling with the strong coupling regime of confining SYM theory in a unifying picture.

On the field theory side it is known that, to some extent, such an interpolating picture exists if we compactify one spatial dimension and consider SYM on ${\mathbb{R}}^{1,2}\times S^1$ \cite{Davies:1999uw}. In this case the non-perturbative physics is much better understood and typically infrared phenomena (such as gaugino condensation) have a semiclassical exhaustive description. It is indeed possible to explicitly write a non-perturbatively generated superpotential that leads to  a mass gap (providing a mass for the ``magnetic«" photons) and gaugino condensation.
  
To investigate SYM theory on the cylinder, we look for the proper modification of the MN background. When one deals with compact directions (as in the cylinder geometry we are considering here) the natural thing to do is T-duality. We then T-dualize the IIB MN solution along one of its flat spatial directions and consider the corresponding type IIA solution. This could be enough to study SYM on the cylinder, but the dilaton still diverges. As it is well known, this is a sign of the opening of the eleventh dimension. We then uplift the solution to 11 dimensions and find a globally well behaved solution. In this set-up SYM theory is the theory living on the worldvolume of $N$ M5-branes that wrap a three cycle with topology $S^2\times S^1$. Their backreaction generates the dual background. 
Being this solution related to the MN one by dualities, it could look as completely equivalent to the latter. But, as we show in this paper, this is not the case and the eleven dimensional solution encodes in a non-trivial way the information that the dual SYM theory has one compact spatial direction. At the level of lower dimensional gauged supergravity, the solution we discuss is related to the six dimensional $F(4)$ gauged supergravity discussed in \cite{Nunez:2001pt}. 


In principle the solution we build in the way sketched above is just valid to describe the infrared of  ${\cal N}=1$ SYM. Perhaps interestingly, we find that  also its UV description (related to NS5-branes in type IIB) corresponds to the same 11-dimensional solution. We have then a unique picture connecting the UV and the IR of the gauge theory in terms of the worldvolume theory of $N$ M5-branes in the background we find. 

On this solution we will perform various gauge theory computations and we will find perfect agreement with expectations. Both for the perturbative and non-perturbative calculations, it will be crucial to use the M5 and M2 worldvolume actions. In particular, we find that in this set-up the theory is naturally formulated in terms of the scalar field dual to the three dimensional vector and such dualization has not to be imposed, as it is usually done, by hand. Such degree of freedom will be related to the ``self-dual'' three form living on the M5 worldvolume (to which the boundary of the M2-branes couples). 

Using the M5 worldvolume theory, we compute the tension of a generic $(p,q)$ string in this cylindrical geometry. Qualitatively (with very small quantitative difference), we get the same result of \cite{Firouzjahi:2006vp}. Thus, the expectation that these objects are good candidates for cosmic strings in the brane inflationary scenario \cite{Kachru:2003sx,Dvali:1998pa} is confirmed also in the case with one compact dimension. Moreover, taking the compact dimension to be of very small size, we deal effectively with three dimensional field theory. We find that for this three dimensional field theory, the formula for the string tension is inherited from its four dimensional cousin without any modification. This does not agree with the near quadratic scaling hypothesis \cite{Ambjorn:1984yu,Lucini:2001nv} for the confining string tension in 2+1 dimensional field theory. It seems rather to agree with the predictions coming from the Maldacena-Nastase solution \cite{Maldacena:2001pb} (see also \cite{Herzog:2002ss}).

The main novelty perhaps is that in this solution new kinds of instantons (the so-called``fractional instantons") have a natural description. They are responsible  for the generation of the non-perturbative superpotential that we compute. 
In the M-theory description, they correspond to Euclidean M2 branes wrapping a 3-cycle. Precisely two zero-modes are left by such configurations. This is the right number to generate a non-perturbative contribution to the superpotential \cite{Witten:1996bn}. 

The paper is organized as follows. In section \ref{secTdual} we write the type IIA supergravity solution relevant to study SYM theory on the cilinder (${\mathbb{R}}^{1,2}\times S^1$). In section \ref{sec11} we uplift the solution to 11 dimensions, finding a more satisfying picture that smoothly interpolates between the UV and the IR of SYM. In this set-up we discuss the gauge-string dictionary. In section \ref{M5} we identify the three dimensional scalar dual to the vector field, relating it to a two-form degree of freedom living on the M5 worldvolume. This identification applies to more general cases than the one considered here. In section \ref{supot} we move to discuss the appearance of fractional instantons. They allow us to compute the non-perturbative superpotential driving gaugino condensation. In section \ref{secpq} we compute the $(p,q)$ string tension. Section \ref{secKS} is dedicated to generalize our finding to the KS solution. We finally write our conclusions and comments on possible speculations. In appendix \ref{appD4} we discuss the general identification of the scalar field living in the flat directions of the M5 worldvolume with topology ${\mathbb{R}}^{1,2}\times S^1\times S^2$. As a concrete example, we recover the perturbative quantum metric of the moduli space of ${\cal N}=4$ three-dimensional SYM theory, the same discussed in \cite{DiVecchia:2001uc} in the type IIA set-up. In appendix \ref{appBPS} we show how the BPS equations of eleven dimensional supergravity are solved by our solution. Appendix \ref{appk} is devoted to a K-symmetry analysis. 

\section{T-dual of the Maldacena-N\'u\~nez solution}  \label{secTdual}

The Maldacena-Nu\~nez background is a solution of the equations of motion of
type IIB supergravity which preserves four supersymmetries. The ten dimensional metric in string frame is:
\be
ds^2_{10}\,=\,g_s \alpha' Ne^{\phi}\,\,\Big[\,
\frac{dx^2_{1,3}}{g_s \alpha' N}\,+\,e^{2h}\,\big(\,d\theta_1^2+\sin^2\theta_1 d\phi_1^2\,\big)\,+\,
d\rho^2\,+\,{1\over 4}\,\sum_{i}(w^i-A^i)^2\,\Big]\,\,,
\label{metric}
\ee
where $\phi$ is the dilaton, $h$ is a function which depends on the dimensionless radial coordinate $\rho$, the
one-forms $A^i$ $(i=1,2,3)$ are
\be
A^1\,=\,-a(\rho) d\theta_1\,,
\,\,\,\,\,\,\,\,\,
A^2\,=\,a(\rho) \sin\theta_1 d\phi_1\,,
\,\,\,\,\,\,\,\,\,
A^3\,=\,- \cos\theta_1 d\phi_1\,,
\label{oneform}
\ee
and the $w^i$'s are  $su(2)$ left-invariant one-forms,
satisfying  $dw^i=-{1\over 2}\,\epsilon_{ijk}\,w^j\wedge w^k$. 
The $w^i$'s parameterize a three-sphere and can be represented in
terms of three angles $\phi_2$, $\theta_2$ and $\psi$:
\be  \label{forms3}
w^1&=& \cos\psi d\theta_2\,+\,\sin\psi\sin\theta_2
d\phi_2\,\,,\rc
w^2&=&-\sin\psi d\theta_2\,+\,\cos\psi\sin\theta_2 
d\phi_2\,\,,\rc
w^3&=&d\psi\,+\,\cos\theta_2 d\phi_2\,\,.
\ee
The angles $\theta_i$, $\phi_i$ and $\psi$ take values in the intervals $\theta_i\in [0,\pi]$, 
$\phi_i\in [0,2\pi)$ and $\psi\in [0,4\pi)$. The functions $a(\rho)$, $h(\rho)$ and the dilaton $\phi$
are the following ones:
\be
a(\rho)&=&{2\rho \over \sinh 2\rho}\,\,,\rc
e^{2h}&=&\rho \coth 2\rho\,-\,{\rho^2\over \sinh^2 2\rho}\,-\,
{1\over 4}\,\,,\rc
e^{-2\phi}&=&e^{-2\phi_0}{2e^h\over \sinh 2\rho}\,\,.
\label{MNsol}
\ee

This solution of type IIB supergravity also includes a Ramond-Ramond three-form $F_{(3)}$
given by
\be
\frac{F_{(3)}}{g_s \alpha' N}\,=\,-{1\over 4}\,\big(\,w^1-A^1\,\big)\wedge 
\big(\,w^2-A^2\,\big)\wedge \big(\,w^3-A^3\,\big)\,+\,{1\over 4}\,\,
\sum_i\,F^i\wedge \big(\,w^i-A^i\,\big)\,\,,
\label{RRthreeform}
\ee
where $F^i$ is the field strength of the su(2) gauge field $A^i$, defined as $
F^i\,=\,dA^i\,+\,{1\over 2}\epsilon_{ijk}\,A^j\wedge A^k$.\\
We perform T-duality along the Minkowski coordinate $x_3$ and call $z$ to the T-dual coordinate. The formulas relating Type IIB and Type IIA theories via T-duality are given in \cite{me}. We summarise them here for convenience. We denote as $G_{\mu \nu}$ the components of the metric in Type IIA and as  $J_{\mu \nu}$ the components of the metric in Type IIB. $C^{(n)}$ denotes a RR n-form potential. Then,
\be
G_{zz}&=&\frac{1}{J_{x_3x_3}}, \rc
G_{\mu \nu}&=&J_{\mu \nu} \quad \mu, \nu \neq x_3 , \rc
C^{(3)}_{z\mu \nu}&=&C^{(2)}_{\mu \nu}, \rc
2 \phi_a&=&\phi_b\equiv \phi ,
\ee
where $\phi_a$ ($\phi_b$) is the dilaton in Type IIA (IIB) supergravity and $C^{(2)}$ verifies the equation $dC^{(2)}= F_{(3)}$. One can integrate this equation obtaining:
\be\label{C2}
&&\frac{C^{(2)}}{g_s \alpha' N}=\frac{1}{4} \Big [ \psi (\sin{ \t_1} d\t_1 \wedge d\phi_1 - \sin {\t_2} d\t_2 \wedge d\phi_2)- \, \rc\rc
&&\cos {\t_1} \cos {\t_2} d\phi_1 \w d\phi_2 
-a (d\t_1 \w \omega^1-\sin{\t_1} d\phi_1 \w \omega^2) \Big ].
\ee
Therefore the Type IIA background we obtain in this way is (in string frame):
\be
ds^2_{10}\,=\,g_s \alpha' Ne^{2 \phi_a}\,\Big[\,\frac{dx^2_{1,2}}{g_s \alpha' N}\,+\,e^{2h}\,\big(\,d\theta_1^2+\sin^2\theta_1 d\phi_1^2\,\big)\,+\,
d\rho^2\,+\,{1\over 4}\,\sum_{i}(w^i-A^i)^2\,\Big] +e^{-2 \phi_a} dz^2 \,\,, \rc
\label{metric1}
\ee
where $\phi_a=\frac{\phi}{2}$ and the RR potential $C^{(3)}=C^{(2)} \w dz$ satisfies the equation
\be
dC^{(3)}=F_{(3)} \w dz=F_{(4)}.
\ee
The coordinate $z$ is periodic with period $2\pi R$ and we choose for convenience $R$ to be the self-dual radius $R=\sqrt{\alpha'}$. We will restore the $R$ dependence when needed.

This solution corresponds to having N D4-branes wrapped along $S^2$ and smeared in the $z$ direction. The Yang-Mills dual theory is a three-dimensional ${\cal N}=2$ supersymmetric field theory with gauge group $SU(N)$. It is obtained via compactification from the six-dimensional theory living on the $D4$ worldvolumes plus the $z$-direction along which the branes are smeared. The (twisted) compactification on the $S^2$ is obviously the same as in the MN case. Along the $S^1$ we can think of a standard compactification. Accordingly, the gauge potential $A_z^a$ ($a=1\ldots N-1$ belonging to the Cartan sub-algebra of $SU(N)$) gives rise to $N-1$ massless scalars. This is an ${\cal N}=1$ four-dimensional theory on the ``cylinder" (${\mathbb R}^{1,2}\times S^1$).

\subsection{Gauge theory analysis}

To investigate the Yang-Mills theory dual to this gravitational background we start making a D4 probe computation. The D4 worldvolume action is:
\begin{equation}\label{D4}
S_{probe}=-\tau_4\int d^5\xi  e^{-\phi_a}\sqrt{-det[G+2\pi\alpha' F]}+\tau_4\,2\pi\alpha'\,\int C_3\wedge F,
\end{equation}
where \beq \tau_4=\frac{1}{(2\pi\alpha')^24\pi^2\sqrt{\alpha'}g_s},\eeq
$F$ is a worldvolume gauge field and all bulk fields are understood to be pullbacks onto the brane worldvolume which is parameterized by $\xi\,=\,\{x^0,x^1,x^2,\Omega_2\}$, where $\Omega_2$ is the volume element of the cycle on which the brane is wrapped \cite{Bertolini:2002yr}:
 \beq\Omega_2\,:\ \ \theta_1=\theta_2,\ \phi_1=2\pi-\phi_2\,\, .\label{cycle}\eeq 
Identifying the compact dimensionless transverse scalar field $b$ via \beq\label{zb}z=2\pi\sqrt{\alpha'}\,b,\eeq 
and neglecting the $\rho$-dependent term (signaling that the no-force condition is not fulfilled in this case), we can expand the action (\ref{D4}) in powers of $\alpha'$. Substituting in it the solution (\ref{C2}-\ref{metric1}), one gets:
\beq
S~=~-\int d^3\xi\frac{1}{g^2_{YM3}}\left[\frac{1}{2\alpha'}\partial_ab\partial^ab+\frac{1}{4}F_{ab}F^{ab}\right]+\frac{1}{2}N\frac{(\psi+a(\rho)\sin\psi)}{4\pi}\int d^3\xi\epsilon^{abc}(\partial_{a}b)F_{bc},\label{vector}\eeq
where: \beq \label{g3}\frac{1}{g^2_{YM3}}~=~\frac{N}{16\pi^2}Y(\rho)~(2\pi\sqrt{\alpha'}),\hspace{1.2cm} \mbox{with}\ \ Y(\rho)=4\rho\tanh\rho.\eeq
If we remember also that, according to \cite{DiVecchia:2002ks}, in the four dimensional theory there is a non-trivial topological term proportional to the $\theta_{YM}$-angle defined as\footnote{Here there is a factor of 2 of difference with respect to \cite{DiVecchia:2002ks} because of the definition of the cycle $\Omega_2$.} \beq \theta_{YM}~=~-N(\psi+a(\rho)\sin\psi),\label{thetaym}\eeq we see that the action (\ref{vector}) corresponds to four dimensional gauge theory compactified on a spatial $S^1$.

A better formulation of such theory is given in terms of the scalar dual to the vector field. To introduce it, we add to the action (\ref{D4}) the term\beq\label{lagrange} -\int\Sigma dF,\eeq 
where the auxiliary field $\Sigma$ acts as a Lagrange multiplier for the Bianchi identity constraint.  Due to the Dirac quantization condition 
\be
q\,=\,\frac{1}{8\pi}\int d^3\xi\epsilon^{abc}\partial_{a}F_{bc}\,\ \in \ {\mathbb Z}\,\, ,
\ee
the auxiliary field $\Sigma$ is periodic, with period\be T_{\Sigma}=\frac{1}{2}\,\, .\ee

A (magnetic) dual description of such theory can be obtained by promoting $\Sigma$ to a dynamical field \cite{Polyakov:1976fu} and integrating out the Abelian field strength $F$ via its equation of motion. One gets:
\beq\label{threed}
S~=~-\frac{1}{2}\int d^3\xi\left[ \frac{1}{g^2_{YM3}\alpha'}\partial_ab\partial^ab~+~g^2_{YM3}\Big(\frac{N(\psi+a(\rho)\sin\psi)}{4\pi}\partial b+\partial\Sigma\Big)^2\right],
\eeq
or using (\ref{thetaym}), one can write (\ref{threed}) as \beq \label{3dual}
S~=~-\frac{1}{2}\int d^3\xi\left[ \frac{2\pi}{g^2_{YM4}R}\partial_ab\partial^ab~+~\frac{g^2_{YM4}}{2\pi R}\Big(\partial\Sigma-\frac{\theta_{YM}}{4\pi}\partial b\Big)^2\right],\eeq
where
\be
\frac{1}{g^2_{YM4}}~=~\frac{N}{16\pi^2}Y(\rho).
\ee
Notice that equation (\ref{3dual}) looks pretty similar to eq. (3.7) of \cite{Seiberg:1996nz} where this theory has first been considered. 
\setcounter{equation}{0}

\section{Eleven dimensional solution dual to Maldacena-N\'u\~ nez}
\label{sec11}

In the previous section we have given a picture of the Yang-Mills theory on the cylinder in the type IIA set-up. One unsatisfactory aspect could be the bad behavior of the dilaton at large values of the radial coordinate. As it is well known, this signals the decompactification of the eleventh dimension of M-theory. Under this perspective, it is quite natural to up-lift the solution to eleven dimensional supergravity.  One gets:
\be  \label{11metric}
\nonumber ds^2_{11}&=& e^{4/3\phi_a}\left[ dx_{1,2}^2+\alpha'g_sN e^{2h}(d\theta_1^2+\sin^2\theta_1 d\phi_1^2)+\alpha'g_sNd\rho^2+\frac{\alpha'g_sN}{4}\sum_i(\omega^i-A^i)^2\right]\\ &&\ \ +\, e^{4/3\phi_a}dy^2~+\ e^{-8/3\phi_a}dz^2\label{11metric},\\ C^{(3)}&=&C^{(2)}\wedge dz,\nonumber\ee 
where $C^{(3)}$ is the magnetic potential under which the $M5$-branes are charged and the new eleventh coordinate $y$ is periodic, with period $2\pi g_s\sqrt{\alpha'}$.
 
This solution corresponds to have $N$ M5 branes wrapping, besides the $S^2$, the $y$ circle. The gauge theory we are describing corresponds thus to the worldvolume theory of such $M5$-branes. We begin noticing an intriguing property of such description.
Starting from eleven dimensions, for small $\rho$ it is fine to go down to IIA along the $y$-circle: it has small radius and the IIA theory is well behaved. We get D4-branes and the proper field theory description we have given in the previous section.

For big $\rho$ instead the radius of the $y$-circle becomes big. If we insist on making the dimensional reduction along that circle, the IIA dilaton diverges. But, as it is clear from eq. (\ref{11metric}), the radius of the $z$-circle becomes small in the large $\rho$ limit: it is now possible to reduce along $z$ and get a well behaved IIA solution in terms of NS5-branes. 
The same happens in the Type IIB MN solution, where for small $\rho$ one has a well behaved solution in terms of D5 branes, while for big $\rho$ the dilaton diverges and one needs to S-dualize the background and get instead NS5 branes. 

We see here as the eleven dimensional picture gives a unifying picture of this phenomenon in terms of a unique theory on the M5-brane worldvolume. Schematically we can summarise the solution in the following table:
\vskip .7cm
\begin{tabular}{cccccr}
 & $\Ccya{\footnotesize{T_1}}$ & & \Ccya{\footnotesize{$\tilde{UP}$}} & &\\ 
\Large{\Cblu{IIB}}&$\hspace{.5cm}\Ccya{\longrightarrow}\hspace{.5cm}$&\Large{\Cblu{IIA}}& $\Ccya{\longrightarrow}$ &\Cblu{\Large{11D}} & \\
\footnotesize{D5} & & \footnotesize{D4}& &\footnotesize{M5} & \\
\\
$\Ccya{\footnotesize{S}\updownarrow}$ & & & &\large{$\Ccya{|||}$} 
& $\hspace{-.1cm}{\mathbb R}^{1,2}\times S^1\times\tilde{S}^1\times CY_3$ \\  
\\
 & $\Ccya{\footnotesize{\tilde{T}_1}}$ & & \Ccya{\footnotesize{UP}} & & \\ 
\Large{\Cblu{IIB}}&$\hspace{.5cm}\Ccya{\longrightarrow}\hspace{.5cm}$&\Large{\Cblu{IIA}}& $\Ccya{\longrightarrow}$ &\Cblu{\Large{11D}} & \\
\footnotesize{NS5} & & \footnotesize{NS5}& &\footnotesize{M5} & \\
\end{tabular}
\vskip .5cm\noindent
We move on now to investigate this wrapped M5 worldvolume theory.

\setcounter{equation}{0}

\section{M5-brane worldvolume theory}  \label{M5}

The analysis we are going to do in this section applies to more general cases than the one under consideration. We perform it in the case that the M5 worldvolume geometry is of the form ${\mathbb{R}}^{1,2}\times S^1\times \Omega^2$, where $\Omega^2$ is a two dimensional compact manifold whose volume element is $V_2d\Omega_2$ ($\int d\Omega_2=4\pi$). $S^1$ is a circle of radius $g_s\sqrt{\alpha'}$. Standard Kaluza-Klein reduction on the internal three dimensional manifold ${\bf{S}}\,=\,S^1\times\Omega^2$ gives rise to the three dimensional gauge theory. To study such theory we need to investigate the M5 worldvolume dynamics. Our main tool will be the (covariant) worldvolume action written by Pasti, Sorokin and Tonin in \cite{Pasti:1997gx}. In such formalism (usually called PST formalism), it is better if the internal manifold contains a factorized circle (or more generally its first Betti number should be different from zero)\footnote{This is related to the fact that the PST scalar is naturally an angular variable.}. 
Even if there are well known problems referred to a possible quantization of the M5-brane action \cite{Witten:1996hc}(problems shared by the  PST formalism with many others), here we are just interested in analyzing the classical equations of motion. To this aim we can safely use the PST formalism.     

The M5 worldvolume PST action is \cite{Pasti:1997gx}:
\be\label{PST}
S&=&T_{M5}\int d^6\xi\left(-\sqrt{-det(g+\tilde{H})}+\frac{\sqrt{-det g}}{4\partial a\cdot\partial a}\partial_ia(\star H)^{ijk}H_{jkl}\partial^la\right)\label{BIPST}\nonumber\\&&\label{CSPST}+
\frac{T_{M5}}{2}\int F\wedge C^{(3)}\, ,\ee
where $g$ is the pullback of the eleven dimensional background metric, $a$ is the PST scalar and the three-form $H$ is defined as
\beq H=F-C^{(3)},\eeq where $F$ is a worldvolume three-form field strength ($F=dA_{(2)}$).
The two-form $\tilde{H}$ is defined as 
\beq
\label{Htilde}\tilde{H}^{ij}~=~\frac{1}{3!\sqrt{-detg}}\frac{1}{\sqrt{-(\partial a)^2}}\epsilon^{ijklmn}\partial_kaH_{lmn},\eeq
and the M5-brane tension is given by \beq \label{TM5}T_{M5}=\frac{1}{(2\pi)^5g_s^2\alpha'^3}\,\, .\eeq

Specifying to our case, we want to implement the KK reduction on the internal manifold ${\bf S}$. We need to do a gauge fixing for the PST scalar. The most natural one is   
 \beq \label{aa}a~=~y,\eeq where $y\in [0,2\pi \sqrt{\alpha'}g_s]$ parameterizes the $S^1$. Consistently with this choice, we consider the following two-form worldvolume potential $A_{(2)}$:\beq \label{AA}\frac{A_{(2)}}{(2\pi)^2g_s}=\alpha'\frac{y}{2\pi g_s}\frac{1}{2} F_{ab}dx^a\wedge dx^b+\alpha'^{\frac{3}{2}}\Sigma d\Omega_2,\eeq   
where the indices $a,\, b$ span the three dimensional Minkowski space ($a,\ b=0,1,2$), $F_{ab}=\partial_aC_b-\partial_bC_a$, $C_a$ is a three dimensional vector and  $\Sigma$ is a dimensionless scalar field depending on the coordinates $x_a$. 
Even if the properties of the quantum M5-brane theory are subtle, it is quite natural to quantize the possible variations of the two-form integrated on the (contractible) two-cycle. One way to see this is to use dualities that relate the two form $A_{(2)}$ to the NS-NS $B_{(2)}$ field. The relevant quantity which is allowed to change by integer units is:
\be
\frac{1}{(2\pi)^3g_s\alpha'^{3/2}}\int A_{(2)}.
\ee 
The field $\Sigma$ is thus naturally periodic. With our normalizations (\ref{AA}) its period is:
\be
T_{\Sigma}\,=\,\frac{1}{2}.\label{TSigma}
\ee
Inserting the ansatz (\ref{AA}) in (\ref{BIPST}) one gets the three dimensional action:
\beq
S=-8\pi^2 g_s \alpha'^{1/2}T_{M5}\left[V_2\int d^{1,2}x\sqrt{-\det(g+\tilde{H})_{ab}}\,+\,\frac{(2\pi)^3g_s\alpha'^{5/2}}{4}\int d^{1,2}x\epsilon^{abc}F_{ab}\partial_c\Sigma\right],
\eeq
and it is straightforward to see that 
\beq
\det(g+\tilde{H})_{ab}=\det (g)_{ab}\left(1+\frac{1}{4}\det(g^{-1})_{ab}\frac{(2\pi)^4g_s^2\alpha'^3}{V_2^2}\partial_c\Sigma\partial^c\Sigma\right)\, . \eeq
Expanding the square root in powers of $\alpha'$ and discarding the constant term, we get the three dimensional flat spacetime action:
\beq\label{mess3}
S=-\left[\frac{1}{2}\frac{2\pi\sqrt{\alpha'} g_s}{V_2}\int d^{1,2}x\partial_c\Sigma\partial^c\Sigma+\frac{1}{2}\int d^{1,2}x\epsilon^{abc}F_{ab}\partial_c\Sigma\right]\, . \label{PST3}
\eeq
Thanks to the second term in (\ref{PST3})  $\Sigma$ is naturally interpreted as the scalar field dual to the vector one. 
We can interprete the vector field $C_a$ as a Lagrange multiplier and to vary with respect to it, enforcing the Bianchi identity constraint on $\Sigma$. Otherwise we can vary with respect to the vector $\partial_a\Sigma$ and get the standard action in terms of the vector field $C_a$. 
As a result, it is quite obvious to relate the volume $V_2$ of the two cycle to the square of the inverse of the three dimensional gauge coupling constant in the following way:
\be
g_{YM3}^2\,=\,\frac{2\pi\sqrt{\alpha'} g_s}{V_2}. 
\ee
It is easy now to recognize (\ref{mess3}) as the standard free Maxwell action in three dimensions.


Notice that the dualization term, which allows us to identify the scalar $\Sigma$ as the dual to the vector field, here it is contained automatically in the worldvolume theory. Such theory is moreover automatically written in terms of the scalar $\Sigma$, the kinetic term for this field appearing in (\ref{PST3}).

As already stressed, these results are quite general for geometries of the type  ${\mathbb{R}}^{1,2}\times S^1\times \Omega^2$. To test this generality, in Appendix A we apply them to the case of ${\cal N}=4$ supersymmetric (8 real supercharges) three dimensional Yang-Mills theory. This theory has a nice geometric description in terms of an Hyper-K\"ahler moduli space (a possible generalization of the Atiya-Hitchin manifold). We refer the reader to \cite{Seiberg:1996nz,Dorey:1997ij} for the details. The explicit gravity dual description has instead been given in \cite{DiVecchia:2001uc} in the type IIA theory. Basing on that paper we discuss the eleven dimensional supergravity dual and show how the proper (perturbative) space emerges naturally in this case. 

\setcounter{equation}{0}

\section{${\cal N}=1$ super Yang-Mills on the cylinder} 
\label{supot}

We are now ready to interpret the eleven dimensional solution given in section \ref{sec11} from the field theory point of view. This is a ${\cal N}=2$ supersymmetric field theory in three dimensions obtained from a circle reduction of ${\cal N}=1$ (pure) super Yang-Mills in four dimensions. Being the $M5$-branes smeared in the $z$-circle, the gauge group is $U(1)^{N-1}$ (the degree of freedom corresponding to the center of mass of the system is decoupled). Applying the analysis we described in the previous section, namely making an M5-probe computation and expanding the result in powers of $\alpha'$, for the {\it i}-th gauge group we get:  
\begin{equation}
S_i=-\frac{1}{2}\int d^3\xi\left[ \frac{2\pi}{g^2_{YM4}R}\partial_ab_i\,\partial^ab_i\,+\, \frac{g^2_{YM4}}{2\pi R}\partial_a\Sigma_i\partial^a\Sigma_i\right]\,\, ,
\end{equation}
where we omit the dualization term and restrict ourselves to the case $\theta_{YM}=0$ (see eq. (\ref{3dual})). The periodic scalars $b_i$ correspond to fluctuations in the $z$ direction and are defined as in (\ref{zb}). Redefining the field $\Sigma_i$ as \be\label{gamma} \gamma_i\,=\,\frac{g^2_{YM}}{2\pi}\Sigma_i,\ee
it is easy to write the action in terms of the holomorphic field \be\label{Phi}\Psi_i\,=\,b_i\,+\,i\gamma_i\ee simply as:
\be 
S_i\,=\,-\frac{\pi}{g^2_{YM}R}\int d^3x\,\partial_a\Psi_i\partial^a\Psi_i\,\, .
\ee
From (\ref{TSigma}, \ref{gamma}) we can read off the period of $\gamma_i$:\be\label{Tgamma}T_{\gamma}\,=\,\frac{g^2_{YM}}{4\pi}.\ee

We look now for some M-brane configuration generating a superpotential. For this to happen one has to check that in presence of such M-branes there are two fermionic zero-modes \cite{Witten:1996bn}.
The superpotential they generate is \cite{Witten:1996bn}:
\be \label{supotM2}W\,\sim\,\mu^3\sum_i{\rm e}^{i\,S_{M_i}},\ee
where $\mu$ is a dimensionfull scale (with inverse length dimension) related to the value of the radial variable at which the computation is made (it is the same at which the actions $S_{M_i}$ are evaluated). 

To have some intuition on what are these configurations, we start noticing that an instanton configuration is an (Euclidean) M2-brane wrapped along $\Omega_2$ (defined in (\ref{cycle})) and the entire $z$ circle. In presence of such configuration, from the index theorem, we expect to have $2N$ fermionic zero-modes. 
Before doing the zero-mode counting, we have to remember that along the $z$-circle there are $N$ $M5$-branes. It is well known that an $M2$-brane can end on an $M5$-one. Analogously, the instantonic $M2$-brane can open itself and end on one of the $N\ M5$s. We are thus led to consider $N$ objects that are more basic than the instantonic $M2$-brane and can be seen as its constituents. These objects are the $N$ $M2$-branes stretching between two consecutive $M5$-branes. 
In this way we get a picture very close to the field theoretical one of the instanton as being composed by more fundamental instanton partons \cite{Belavin:1979fb}, the so-called "fractional instantons'' (our open $M2$-branes). 

For the Kappa-symmetry analysis of this kind of configurations, we refer the reader to appendix C. There, we show that half of the supersymmetries of the background are preserved (in the proper limit) by such M2-branes. This implies that two (real) supersymmetries are broken and, consequently, there are two fermionic zero modes in this background: they are the goldstinos of the broken supersymmetries. The equations of motion for the fermionic fluctuations (up to second order in fermions) in an arbitrary bosonic background have been written in \cite{Mart}. As a direct inspection of such equations shows, the presence of other zero modes is unlikely. Therefore, we assume that in this background there are just the two goldstinos zero modes we discussed here.

\subsection{The non-perturbative superpotential}

To proceed further and write explicitly the superpotential generated by these M2-branes via the formula (\ref{supotM2}), we need the form of their worldvolume action. As we are considering open M2-branes stretching between two M5s, we have to pay attention to the fact that on the $M5$-brane worldvolume, the boundary of an $M2$-brane (a string) sources a potential. This is 
precisely the $A_{(2)}$ $M5$ worldvolume two-form potential. The coupling of open $M2$ to it is easily evaluated \cite{Strominger:1995ac} (perhaps the best way of seeing it is by requiring gauge invariance for the $C^{(3)}$ potential). The resulting (open) $M2$ worldvolume action is:
\be\label{M2}
iS_{M2_i}\,=\,-T_{M2}\int d^3\xi\sqrt{det\, g}\,+\, i\,T_{M2}\int\left( C^{(3)}-F\right),
\ee
where $$T_{M2}\,=\,\frac{1}{(2\pi)^2g_s\alpha'^{3/2}}.$$

Out of the $N$ M5-branes we need to decouple the center of mass. To this aim, we consider one (non dynamical) $M5$-brane fixed at $z=0$. For the {\it i}-th $M2$-brane (extending between the $({\it{i}}-1)$ and the {\it i} $M5$ ones), we can evaluate the action (\ref{M2}):
 \be
 iS_{M2_i}\,=\,-\frac{8\pi^2}{g^2_{YM}}\left[(b_i-b_{i-1})\,+\,i\,\frac{g^2_{YM}}{2\pi}(\Sigma_i-\Sigma_{i-1})\right]\,=\,-\frac{8\pi^2}{g^2_{YM}}\left(\Psi_i-\Psi_{i-1}\right)\,=\,-\frac{8\pi^2}{g^2_{YM}}(\Delta\Psi)_i \,\, , \rc
\ee
 where we make the computation at $\theta_{YM}=0$ and the definitions (\ref{zb},\ref{AA},\ref{Phi}) are used. We must pay special attention to the {\it N}-th $M2$-brane, the one extending between the $N-1$ M5-brane and the non dynamical one. In this case its action is not independent of the others, but it is given by the difference between the instantonic one (the one corresponding to the $M2$ extending along the entire circle $z$) and all the others $N-1$. We call it Kaluza-Klein monopole (as it has been named the analogous configuration in field theory in \cite{Davies:1999uw}). We easily compute its action\footnote{We call it $S_{KK}$ but it has not to be confused with the action for the KK-modes coming from the $S^2$ worldvolume compactification.} and  find perfect agreement with field theory (see eq. (2.10) of \cite{Davies:1999uw}):
 \be
 S_{KK}\,=\,-\frac{8\pi^2}{g^2_{YM}}\frac{R}{\sqrt{\alpha'}}\,+\, \sum_{i=1}^{N-1}\frac{8\pi^2}{g^2_{YM}}(\Delta\Psi)_i.
 \ee
 
Putting all together and redefining $(\Delta\Psi)_i$ as $\Phi_i$, we get the superpotential (\ref{supotM2}):
 \be\label{M2supot}
 W\,=\,M^3\left(\sum_{i=1}^{N-1}{\rm e}^{-\frac{8\pi^2}{g^2_{YM}}\Phi_i}+{\rm e}^{-\frac{8\pi^2}{g^2_{YM}}\frac{R}{\sqrt\alpha'}+\sum_{i=1}^{N-1}\frac{8\pi^2}{g^2_{YM}}\Phi_i}\right).
 \ee
This nicely reproduces the field theory one \cite{Davies:1999uw}. By extremizing it we get the M-branes equilibrium configurations:
\be    \label{vacua}
&&\langle\Phi_i\rangle\,=\,\frac{R}{N\sqrt{\alpha'}}\,+\,i\,\frac{g^2}{4\pi}\,\frac{ k}{N} \qquad \qquad k \,\, \in \,\, {\mathbb Z} \,\, ,\rc
&& \langle W \rangle\,=\, N\Lambda^3,
\ee
where $k$ is defined modulo N (see (\ref{Tgamma})) and labels the $N$ vacua resulting from the breaking of the ${\mathbb Z}_{2N}$ symmetry. These vacua are related to the ones of gaugino condensation. Domain walls naturally follow.
Via this superpotential, the magnetic photons do get a mass. This is a signal of  confinement. 

\subsection{$U(1)_R$ Anomaly}

Special attention must be paid to analyze the $U(1)_R$ symmetry of the model. This is a proper R-symmetry under which also the $\theta$'s fermionic integration variables transform. It acts in the following way:
\be
\theta\,\to\,{\rm e}^{i\alpha}\theta\hspace{0.1cm},\hspace{0.9cm}\chi_i\,\to\,{\rm e}^{i\alpha}\chi_i,\hspace{0.1cm},\hspace{0.9cm}\bar{\chi}_i\,\to\,{\rm e}^{-i\alpha}\bar{\chi}_i,
\ee
where $\chi_i$ are the fermions in the $U(1)_i$ vector multiplet.

In standard ${\cal N}=2$ three dimensional theory this symmetry is not anomalous, as it is a subgroup of a simple group, $SU(2)_R$. 
The way  the symmetry is not violated at low energy is assigning $U(1)_R$ charge transformation to the magnetic photons \cite{Affleck:1982as}:
\be
\gamma_i\,\to\,\gamma_i\,-\,\frac{g^2}{4\pi^2}\alpha\hspace{0.4cm}{\mbox{ for all $i$}}
\,\, .\ee 

However, we are considering a slightly 
different theory, {\it i.e.} the one obtained by Kaluza-Klein compactification on a circle. In our case  the $SU(2)_R$ is explicitly broken by the circle and the $U(1)_R$ symmetry is anomalous.
 As it is clear from its membrane origin we have just described, the way the theory knows about the extra compact dimension (responsible for the anomaly) is via the Kaluza-Klein monopole. This is the non-perturbative configuration generating the last term in the superpotential (\ref{M2supot}). If we continue to pretend that the superpotential (\ref{M2supot}) has charge two under $U(1)_R$, this term causes the expected anomaly: $U(1)_R$ is no longer a symmetry but just its subgroup ${\mathbb Z}_{2N}$ is preserved. Notice that the first term in  (\ref{M2supot}) has the proper charge without implying any anomaly, being the corresponding monopoles also present in the purely flat case.
 Eventually, in the vacua (\ref{vacua}), the photons condense spontaneously breaking the ${\mathbb Z}_{2N}$ symmetry to ${\mathbb Z}_2$.

 \begin{section}{Dual configurations to Effective Strings}
 \label{secpq}
 
We move now to compute the tension of a generic $(p,q)$ string. By $(p,q)$ string we mean an extended object with $p$ units of fundamental string charge and $q$ units of D1-brane charge. The field theoretic interpretation of such object, which extends in two of the Minkowski directions, is clear in the $q=0$ case \cite{Herzog:2001fq}: it is the confining $p$-sting (the string joining $p$ quarks to $p$ anti-quarks). In the $p=0$ case, its interpretation is clear in the KS solution, where it corresponds to an axionic string \cite{Gubser:2004qj}. In the MN background such BPS object exists but its field theory interpretation is not well understood (no axionics strings are supposed to be present in this case \cite{Gubser:2004qj,Caceres:2005yx}). A first attempt to understand their origin could be to apply the formalism developed in \cite{Gursoy:2005cn} to see if they are related to a pure super Yang-Mills effect or to a Kaluza-Klein one (we are thinking about the KK modes of the $S^2$). In the generic $(p,q)$ case these objects, once the KS solution is embedded in a cosmological (inflationary) scenario \cite{Dvali:1998pa,Kachru:2003sx}, can look like cosmic strings \cite{Copeland:2003bj}, whose range of tension is compatible with the current observational bounds \cite{Shandera:2006ax}.

To compute the tension of a $(p,q)$ string, the relevant geometry is the one corresponding to the deep IR of the field theory, {\it i.e.} at $\rho = 0$.
Such geometry is  ${\mathbb R}^{1,2}\times S^1 \times S^3\times \tilde{S}^1$ with $N$ units of $G$-form flux ($G=dC_{3}$) through $S^1\times S^3$. We call $S^1$ the circle paremeterized by $z$ and $\tilde{S}^1$ the one paremeterized by $y$. 
 
The M-theory description of a $(p,q)$-string is given in terms of a bound state of $p$ M2-branes extending along the coordinates $x^0,\, x^1,\, y$ and other $q$ M2-branes extending along $x^0,\, x^1,\, z$.
Also here the Myers effect takes place for the $p$ M2-branes and they blow up into a M5-brane with $p$ units of worldvolume flux through $S^2\times S^1$ ($S^2$ is the two-sphere inside $S^3$). In this set-up, the presence of the other $q$ M2-branes corresponds to a (quantized) electric flux in the $x^0,\, x^1,\, z$ directions.
We compute now the tension of this object by using the classical action plus a quantization condition on the worldvolume fluxes. This computation is clearly related by a chain of dualities to the one made in \cite{Firouzjahi:2006vp}.

The resulting M5-brane is parameterized by the following set of coordinates:
\beq
\xi^\mu(\mu=0,\ldots,5)= (x_0, x_1, y, z, S^2).
\eeq
The embedding we are interested in is at $\rho=0$ where the remaining scalars do not play any role, except for the $S^3$ polar angle (call it $\psi$) that we assume to depend on the flat coordinate $x_1$,  {\it i.e.} $\psi=\psi(x_1)$. 
Let $\psi, \tilde{\theta}$ and $\tilde{\phi}$ be the coordinates which parameterize the 3-sphere\footnote{The M5-brane extends along the $S^2$ parameterized by $\tilde{\theta}$ and $\tilde{\phi}$.}, with $\psi$ the polar angle $(0< \psi \leq \pi)$. The $S^3$ line element $d\Omega^2_3$ can be decomposed as:
\beq
d\Omega^2_3\,=\,d\psi^2\,+\,\sin^2 {\psi}\, d\Omega^2_2\,\, .
\eeq
In terms of this new set of coordinates, the three-form $C^{(3)}$(see eqs. (\ref{C2}) and (\ref{11metric})) is:
\beq
\frac{C^{(3)}}{\alpha'g_sN}\,=\,C(\psi)\, d\Omega_2 \wedge dz\,\, ,
\eeq
where $C(\psi)$ is
\beq
C(\psi)\,=\,\psi-\frac{1}{2}\sin (2\psi).
\eeq

In order to capture flux and properly describe the Myers effect we switch on a worldvolume two-form $A_{(2)}$ such that
\beq
F\,=\,dA_{(2)}\,=\,F_{x_0 x_1 z} dx_0 \wedge dx_1 \wedge dz \,+\, F_{z\Omega_2} dz \wedge d\Omega_2 \,\, .
\eeq

As discussed in section \ref{M5}, the general expression for the action of a probe M5-brane in the PST formalism is
\be
&&S\,=\,T_{M5}\int d^6\xi\left(-\sqrt{-\det(g+\tilde{H})}+\frac{\sqrt{-\det g}}{4\partial a\cdot\partial a}\partial_ia(\star H)^{ijk}H_{jkl}\partial^la\right)\,+\, \rc
&&+\,
T_{M5}\int\left(\frac{1}{2}F\wedge C^{(3)}\right)\,\, .
\ee
Analogously to what has been done in the computation of section \ref{M5}, we make the gauge choice $a=y$. The lagrangian density for this probe becomes:
 \be  \label{lagran}
 {\cal L}\,&=&\,-T_{M5}(\alpha'g_sN)e^{2\phi_0}\sin{\tilde{\theta}}  \sqrt{ 1+\alpha' g_s N (\partial_{x_1}\psi)^2 -F^2_{x_0 x_1 z} } \sqrt{ \sin^4{\psi}+\left(\frac{F_{z\Omega_2}}{\alpha' g_s N}-C(\psi)\right)^2 } \,\, . \rc\rc
 \ee
 
The quantization condition that accounts for the Myers effect is:
 \beq\label{p}
 \int_{S^1\times S^2} F_{z\Omega_2} \,=\, \frac{2\pi p}{T_{M2}}\,\, , \qquad \qquad p\,\,\, \in \,\,\, {\mathbb Z}\,\, .
 \eeq
Now we need the quantization condition for the remaining component of the worldvolume flux, $F_{x_0x_1z}$. It is clear from eq. (\ref{lagran}) that
 \beq
 \frac{\partial{\cal L}}{\partial F_{x_0x_1z}}\,=\,\tr{const}\,\, ,
 \eeq
 and the constant can be determined by following the procedure explained in \cite{you}. This means that the quantization condition amounts to
 \beq
 \int_{\tilde{S}^1\times S^2} d\Omega_2 dy \frac{\p {\cal L}}{\p F_{x_0x_1z}}\,=\,q\, T_{M2} \quad , \quad q \; \; \in \; \;{\mathbb Z} \,\, .
\eeq
This condition can be rewritten as:
\beq \label{elec}
F_{x_0x_1z}\,=\, \sqrt{\frac{1\,+\,\alpha'g_sN(\partial_{x_1}\psi)^2}{\sin^4{\psi}\,+\,C^2_p(\t)\, + \, e^{-4\phi_0} \, e(\frac{\pi q}{g_sN})^2}}\,\, e^{-2\phi_0}\, \left(\frac{\pi q}{g_sN}\right)\,\, ,
\eeq
 where we have used (\ref{p}) and defined
 \beq
C_p(\psi)=C(\psi)-\frac{\pi}{N}p \,\, .
 \eeq
 
The next step is to find the minimal energy configurations. By performing a Legendre transformation, the hamiltonian can be written as:
\beq 
\hspace{-.9cm}H\,=\, \int_{S^2\times S^1\times \tilde{S}^1} d\Omega_2\, dy\, dz  \int dx_1 \Big [ F_{x_0x_1z}\frac{\p {\cal L}}{\p F_{x_0x_1z}}\,-\,{\cal L} \Big ] \,\, ,
\eeq
and after some calculations, using the quantization conditions (\ref{p}, \ref{elec}), one can see that the hamiltonian takes the simple form
\beq \label{ener}
H\,=\,2\,g_s\,\sqrt{\alpha'}\,N T_{M2}e^{2\phi_0} \int dx_1 \sqrt{1\,+\, \alpha'g_sN(\partial_{x_1}\psi)^2} \sqrt{e^{-4\phi_0}\left(\frac{\pi q}{g_sN}\right)^2\,+\,\sin^4{\psi}\,+\,C^2_p(\psi)}\,\, .
\eeq
The constant $\psi$ configurations that minimize the energy are given by:
\beq
\left(\frac{\p H}{\p \psi} \right) _{\psi=\tr{const}}\,\sim\,\sin{\psi} \cos{\psi}\,+\,C_p(\psi) =0.
\eeq
Defining the function
\beq \label{del}
\Delta(\psi)\,=\,\sin{\psi} \cos{\psi}\,+\,C_p(\psi),
\eeq
the energy is minimized for $\psi=\bar{\psi}_p$ such that
\beq
\Delta(\bar{\psi}_p)\,=\,0 \,\,  ,
\eeq
that is
 \beq
 \bar{\psi}_p\,=\,\pi \frac{p}{N} , \quad 0\leq p < N.
 \eeq
 The energy of these configurations is (from eq. (\ref{ener}))
 \beq
 H_{p,q}\,=\,\int dx_1 T_{p,q}  \,\, , \rc\rc\eeq
 with \beq T_{p,q}\,=\,2\,g_s\,\sqrt{\alpha'}\,N T_{M2}e^{2\phi_0} \sqrt{e^{-4\phi_0}\left(\frac{\pi q}{g_s N}\right)^2\,+\,\sin^2{\bar{\psi}_p}}\,\, 
  \eeq
being the tension of the $(p,q)$-string we were looking for.

\setcounter{equation}{0}

 \section{The conifold and SYM on the cylinder}  \label{secKS}

In the previous sections we have studied ${\cal N}=1$ SYM theory on the cylinder starting from the MN supergravity solution.
We want to show now that the same could have been done for the Klebanov-Strassler solution \cite{Klebanov:2000hb}. 
For the gauge theory dual, we assume that the deep infrared of that solution describes pure ${\cal N}=1$ super Yang-Mills theory with a single $SU(M)$ gauge group. Even if this is the case only in the limit of small 't Hooft coupling \cite{Klebanov:2000hb, Gubser:2004qj}, which is the opposite limit of supergravity, there are convincing arguments stating that the physics is quite similar also in the (large 't Hooft coupling) supergravity limit \cite{Strassler:2005qs}.

Considering the KS solution and making  a T-duality along one spatial flat direction (call it $z$) transverse to the cone, we end up in the type IIA geometry corresponding to having M fractional D2-branes smeared along the $z$ circle.  In this case there are no specific problems with the dilaton and the solution can be conveniently studied directly in the type IIA set-up. It is:
\be
&&ds^2\,=\,h^{-1/2} \Big ( -dx^2_0 \,+\,dx^2_1\,+\,dx^2_2 \Big ) \,+\,h^{1/2} \Big ( dz^2\,+\,ds^2_6 \Big )\,\, , \rc
&&e^{2\phi}\,=\,h^{1/2}\,\, , \rc
&&C_{(3)}\,=\,\frac{h^{-1}}{g_s} dx_0 \wedge dx_1 \wedge dx_2 \,+\, C_{(2)} \wedge dz \,\, , \rc
&&B_2 \,\, ,\label{KSIIA}
\ee
where $ds^2_6$ is the metric of the deformed conifold, $C_{(2)}$ is the type IIB RR potential  satisfying $dC_{(2)}=F_{(3)}$ and all the forms and functions ($B_2$, $F_{(3)}$ and $h(\tau)$) are the ones in the KS solution \cite{Klebanov:2000hb}.

For the gauge theory analysis of the coupling constant one would need to know the precise form of the worldvolume action of fractional D$p$-branes in the conifold geometry. Lacking this knowledge, the best one can do is to borrow the known ${\cal N}=2$ orbifold fractional branes action. This assumption is sensible because the conifold theory can be related to the ${\cal N}=2$ orbifold one, as it has been first derived in \cite{Klebanov:1998hh}\footnote{See \cite{Merlatti:2005cj} for a different way of connecting the two theories.}.

Accordingly, we assume that a generic fractional D$p$-brane in the conifold geometry is described by the natural generalization of the worldvolume action of the orbifold \cite{Douglas:1996xg,Merlatti:2000ne}. This assumption works in many cases. As first noted in \cite{Klebanov:2000hb}, in the $p=3$ case, it gives the expected result for the coupling constant. It reproduces also the right $U(1)_R$ anomaly. As a further check, one can compute the instantonic action in the type IIB original solution. Identifying the instanton with a fractional D(-1)-brane \cite{Klebanov:2000hb}, one gets the expected result.

In the T-dual type IIA solution (\ref{KSIIA}) the instanton becomes a fractional D0-brane extending in the $z$ direction. 
Analogously to what happens in the MN solution, as described in section \ref{supot}, in this cylindrical case the instanton decomposes in instanton partons (the open fractional D0-branes) suspending between two adjacent fractional D2-branes. In the background of each of those ``fractional instantons" there are again two fermionic 0-modes (the goldstinos of the broken supersymmetries): they generate a superpotential of the form (\ref{supotM2}). Now the action in the exponent is the action of the open fractional D0-brane. To write explicitly the superpotential, we need to consider that the boundary of a fractional D0-brane sources a scalar potential on the fractional D2-brane.
The minimal coupling of the RR $1$-form potential to the fractional D$0$-brane is
\beq
\label{RRminimal}
\int_{\Sigma_{z}}C_{1}(1+b),
\eeq
where $b$ is the scalar field resulting from the integral of the NS-NS $B_2$ field on the two cycle of the conifold geometry.
We see that
under a gauge transformation $C_{1}\to C_{1}+d\Lambda$, (\ref{RRminimal}) is not invariant if the line $\Sigma_z$ has a boundary, like the open D$0$-branes  we are considering. To preserve gauge invariance one must introduce the coupling of the boundary of the D$0$-brane with the worldvolume scalar potential of the host D$2$-brane. In this case the boundary of the D0 couples to a worldvolume scalar ($\Sigma$) on the D2: (\ref{RRminimal}) is replaced by
\beq
\label{RRnonminimal}
\int_{\Sigma_{z}}(C_{1}-d\Sigma)(1+b).
\eeq
Gauge invariance is therefore restored if the gauge transformation of $C_1$ is accompanied by $\Sigma\to\Sigma+\Lambda$. This scalar field $\Sigma$ is naturally interpreted as the scalar field dual to the vector one on the D3 worldvolume.

Considering (\ref{supotM2}) in the case of fractional D0-branes, one gets the following superpotential:
\beq
W\,=\, M^3\left(\sum_{i=1}^{N-1}{\rm e}^{-\frac{8\pi^2}{g_{YM}^2}\Phi_i}+{\rm e}^{-\frac{8\pi^2}{g^2_{YM}}\frac{R}{\sqrt\alpha'}+\sum_{i=1}^{N-1}\frac{8\pi^2}{g_{YM}^2}\Phi_i}\right)\,
,\eeq
 in perfect agreement with field theory and the MN case of the previous section. 

\setcounter{equation}{0}
\section{Conclusions}

In this paper we have investigated the stringy dual description of ${\cal N}=1$ SYM theory on the cylinder (${\mathbb R}^{1,2}\times S^1$). This model is interesting because it contains a rich non-perturbative dynamics that can be succesfully described. This makes it appealing for cosmological applications as well.

To describe its UV properties, we have derived an action written in terms of a complex scalar whose imaginary part is dual to the three dimensional vector field. This is better done in eleven dimensions and, as we show, it is a quite general feature of three dimensional gauge theories with a gravity dual. 
After the identification of the field theory monopole configurations (the four-dimensional ``fractional instantons"), we have derived the form of the non-perturbatively generated superpotential. This superpotential is responsible for the generation of a mass gap, the breaking of chiral symmetry and the appearance of domain walls. 
We have moved then to compute the $(p,q)$-string tension. We have got a formula that is also valid for the Kaluza-Klein reduced three dimensional theory. 
 
The cylindrical geometry we consider may also be relevant in the KKLT set-up \cite{Kachru:2003aw}. More specifically, the fractional instantons we find can source the second term in the superpotential
\beq
W\,=\, W_0\, +\, A e^{i a \rho},
\eeq
 where the first term is generated by fluxes, $A$ and $a$ are constants and $\rho$ is a volume modulus. This is crucial for the issue of stabilizing the K\"ahler moduli.
 Moreover, in the scenario of \cite{Kachru:2003sx}, the $(p,q)$ strings we described are good candidates to be cosmological strings.

As future developments, one can try to face unsolved problems using the eleven dimensional background we have presented here. For example, domain walls seem to have a better description in this geometry (this is not surprising as they are very sensitive to the topology of spacetime). In particular, one could try to find the proper M5-brane configuration dual to them and apply the formalism of section \ref{secpq} to compute their tension. One could also try to follow the line of
\cite{Casero:2006pt} and see how the inclusion of flavors modifies the geometry. In this geometry, it should be possible to derive the superpotential corresponding to the Affleck-Dine-Seiberg \cite{Affleck:1983mk} one in the purely flat four dimensional theory.

\medskip
\section*{Acknowledgments}
\medskip 
We are grateful to D. Are\'an, J. D. Edelstein, Javier M\'as, C. N\'u\~nez and Alfonso V. Ramallo for
comments and discussions. This work was partially supported by MCyT and FEDER under grant FPA2005-00188, by Xunta de Galicia (Conseller\'\i a de Educaci\'on and grant PGIDIT06PXIB206185PR) and by the EC Commission under grant  MRTN-CT-2004-005104.

\appendix

\setcounter{equation}{0}
\section{Wrapped D4-branes}
\label{appD4}

To test the generality of the results of section \ref{M5}, we apply our method to another case of branes wrapped on a cycle, namely the case of D4-branes wrapped on a two cycle. The geometry of the solution is the one discussed in \cite{DiVecchia:2001uc}. They studied the solution in the IIA context, where a necessary tool to find the proper Hyper-K\"ahler moduli space is to dualize by hand the three dimensional vector field into a scalar (as it is usual in these cases). We want to see if this dual scalar field emerges again naturally from the degrees of freedom of the M5-brane. In eleven dimensions, the metric of the solution in \cite{DiVecchia:2001uc} becomes\footnote{We refer the reader to \cite{DiVecchia:2001uc} for the notation adopted here.}:
\be
ds^2_{11}&=&e^{4/3\Phi}\left[dx_{1,2}^2+ZR_0^2(d\tilde{\theta}^2+\sin^2\tilde{\theta}d\tilde{\phi}^2)+ dy^2\right] +\\ && e^{-8/3\Phi}\left[dr^2+r^2(d\theta^2+\sin^2\theta d\phi^2)+\frac{1}{Z}\left(d\sigma^2+\sigma^2(d\psi+\cos\tilde{\theta} d\tilde{\phi})^2\right)\right]\nonumber\,\, .
\ee
It is easy to see that probing the solution in a supersymmetric way requires to restrict to the $\sigma=0$ subspace of this solution. On this subspace the RR three form potential reads:
\beq
C_3=\frac{R^3_A}{8}\cos\theta\sin\tilde{\theta}d\tilde{\theta}\wedge d\tilde{\phi}\wedge d\phi \,\, .
\eeq
The solution is given in terms of the following parameter and functions (we write here their expression at $\sigma=0$):
\beq
e^{\Phi}=H(r)^{-1/4}\, , \hspace{1cm}Z=\left(1-\frac{R_A^3}{8R_0^2}\frac{1}{r}\right)\, , \hspace{1cm}R_A=2\sqrt{\alpha'}(\pi g_s N)^{1/3}\,\, .
\eeq
This geometry corresponds to M5-branes extending along the three dimensional Minkowski spacetime and wrapping the compact cycle parameterized by $\tilde{\theta},\ \tilde{\phi},\ y$. To probe this geometry with a similar M5-brane, we make the following ansatz for the worldvolume two form:
\be
A_{(2)}=4\pi^2g_s\sqrt{\alpha'}\alpha'\Sigma d\tilde{\theta}\wedge\sin\tilde{\theta} d\tilde{\phi}\,\, ,
\ee
where the fields are normalized as in (\ref{AA}) and we omit the dualization term. Then, at leading order in $\alpha'$, we have
\be
\tilde{H}_{ab}\,=\,i\frac{4\pi^2g_s\alpha'^{3/2 }}{ZR_0^2}e^{-2/3\Phi}\epsilon_{abc}\left(\partial^{c}\Sigma-\frac{N}{4\pi}\cos\theta\partial^{c}\phi\right),
\ee
where we define, consistently with \cite{DiVecchia:2001uc}, the dual three dimensional Yang-Mills coupling constant as
\be 
\frac{1}{g^2_{YM}}=\frac{R_0^2}{2\pi g_s\sqrt{\alpha'}}Z(r).
\ee
We can now expand the M5-brane action at quadratic order in $\alpha'$ and integrate it along the compact directions. We get:
\be
S_{M5}=\int d^3x \left[\frac{1}{g^2_{YM}}\left(d\mu^2+\mu^2(d\theta^2+\sin^2\theta d\phi^2)\right)+g^2_{YM}\left(d\Sigma-\frac{N}{4\pi}\cos\theta d\phi\right)^2\right],
\ee
where we have used the obvious radius-energy relation $r=2\pi\alpha'\mu$. This is the expected perturbative quantum metric on the moduli space (the Taub-NUT metric).

\setcounter{equation}{0}
 \section{Supergravity solution in eleven dimensions}
 \label{appBPS}
 
 For the metric given in equation (\ref{11metric}), let us consider the orthogonal frame
 \be  \label{frame}
 &&e^{x^i}\,=\,e^{2/3\phi_a}dx^i \,\, ,
\qquad (i=0,1,2) \qquad  \; e^{\rho}\,=\,e^{2/3\phi_a}(\alpha'g_sN)^{1/2}d\rho \,\, , \rc\rc
&&e^{y}\,=\,e^{2/3\phi_a}dy \,\, , \qquad \qquad \qquad \qquad \qquad e^{z}\,=\,e^{-4/3\phi_a}dz \,\, , \rc\rc
&&e^1\,=\, e^{2/3\phi_a}(\alpha'g_sN)^{1/2}e^{h}d\theta_1 \,\, , \qquad \qquad \; \,\,  e^2\,=\, e^{2/3\phi_a}(\alpha'g_sN)^{1/2}e^{h} \sin{\theta_1} d\phi_1 \,\, , \rc\rc
&&e^{\hat{i}}\,=\, e^{2/3\phi_a}\left(\frac{\alpha'g_sN}{4}\right)^{1/2}(\omega^i\,-\,A^i) \,\, , \qquad (i=1,2,3)\,\, .
 \ee
 
We want to find the explicit form of the Killing spinor and the projections it has to satisfy to solve the eleven dimensional BPS equation. Such equation can be written as follows:
 \beq\label{BPS11}
 \nabla_M \epsilon \,+\, \frac{1}{288} \Big ( \Gamma_{M}^{\,\,NPQR}G_{NPQR}\,-\,8\Gamma^{PQR}G_{MPQR} \Big )\epsilon\,=\,0\,\, ,
 \eeq
 where $M,N,P\ldots$ are curved indices in eleven dimensions, $\nabla$ is the covariant derivative\beq\nabla_M\,=\,\partial_M\,+\,\frac{1}{4}w_M^{ab}\Gamma_{ab}\, ,\eeq $w_M^{ab}$ are the components of the spin connection ($a,b,\ldots$ are flat indices)  and $G$ stands for the four form field strength, $G\,=\,dC_{(3)}$. In the above  frame (\ref{frame}), $G$ is written as
 \be
G\,&=&\, e^{-2/3 \phi_a}(\alpha'g_sN)^{-1/2} \Big [  -2 e^{\hat{1}} \wedge e^{\hat{2}} \wedge e^{\hat{3}} \,-\,\frac{a'}{2}e^{-h}e^{\rho} \wedge e^{1} \wedge e^{\hat{1}}\,+\,\frac{a'}{2}e^{-h}e^{\rho} \wedge e^{2} \wedge e^{\hat{2}}\,+\,\rc\rc
&+&\, \frac{(1-a^2)}{2}e^{-2h} e^{1} \wedge e^{2} \wedge e^{\hat{3}}\Big ] \wedge e^{z}\,\, .
 \ee
With straightforward computations it is possible to see that the Killing spinor satisfying the BPS equation (\ref{BPS11}) is: 
 \beq
 \epsilon\,=\,e^{-\alpha/2\Gamma_{2\hat{2}}}e^{\phi_a/3}\eta \,\, ,
 \eeq
 where $\alpha$ is the angle determined in \cite{Nunez:2003cf}:
 \be
cos \alpha\,=\, \frac{e^{2h}-\frac{1}{4}(a^2-1)}{\rho}   \,\, , \qquad \qquad \sin \alpha\,=\,-\frac{ae^{h}}{\rho}\,\, ,
 \ee
  and $\eta$ is a constant spinor satisfying the projections:
 \beq  \label{projections}
 \Gamma_{12}\eta\,=\,\Gamma_{\hat{1}\hat{2}}\eta \,\, , \qquad \Gamma_{\rho\hat{1}\hat{2}\hat{3}}\eta\,=\,\eta \,\, , \qquad \Gamma_{z}\eta\,=\,\eta \,\, .
 \eeq
 The number of real components of the Killing spinor preserved by this solution is four. Accordingly, the dual field theory is ${\cal N}=1$ SYM in four dimensions.
 
 \setcounter{equation}{0}
 \begin{section}{Kappa symmetry analysis}
 \label{appk}
 
 We look for a supersymmetric configuration of an M2-brane in the background described above. Being the analysis pretty close to the one in \cite{Nunez:2003cf}, we refer the interested reader to that paper for generalities about the Kappa symmetry analysis and its application to this case.
 
Consider an Euclidean M2-brane wrapping the 3-cycle $S^2 \times S^1$, where the $S^2$ is the one defined in (\ref{cycle}) and the $S^1$ is parameterized by $z$. 
Without worldvolume gauge fields, the Kappa symmetry matrix acting on the Killing spinor reduces to:
 \beq
 \Gamma_{\kappa}\,\epsilon=\,\frac{1}{[e^{2h}\,+\,\frac{1}{4}(a-1)^2 ]} \Big (  [e^{2h}\,-\,\frac{1}{4}(a-1)^2 ] \Gamma_{12}\,-\,e^{h}(a-1) \Gamma_{1\hat{2}}\Big )\epsilon \,\, ,
 \eeq
 where $\epsilon$ is the Killing spinor satisfying the projections (\ref{projections}).
The case we are studing is a particular case of the general embedding proposed in section 3 of \cite{Nunez:2003cf}, where a D5-brane in type IIB supergravity is considered. Although we are now dealing with an M2-brane in eleven dimensional supergravity, we can apply the conclusion of their analysis to our case and say that this configuration satisfies Kappa symmetry only asymptotically ($\rho \to \infty$ limit). The only difference is that we have to impose an additional projection (commuting with the ones in (\ref{projections})):
 \beq  \label{kappa}
 \Gamma_{12}\,\epsilon=\,\epsilon \,\, .
 \eeq
This further projection breaks half of the supersymmetries of the background: two real supercharges are preserved.
 
 \end{section}

  \end{section}

\end{document}